\documentclass[12pt,a4paper,preprint,superscriptaddress]{revtex4-1}

\usepackage[caption=false]{subfig}
\usepackage{float}
\usepackage{amsmath,amssymb}
\usepackage[utf8]{inputenc}
\usepackage{graphicx}
\usepackage[english]{babel}
\makeatletter\AtBeginDocument{\let\@elt\relax}\makeatother

\begin{document}


\title{Molecular Dynamics Study on the Role of Ar Ions in the Sputter Deposition of Al Thin Films}
\date{\today}
\author{Tobias Gergs}
\affiliation{Chair of Applied Electrodynamics and Plasma Technology, Department of Electrical Engineering and Information Science, Ruhr University Bochum, 44801 Bochum, Germany}
\email[]{tobias.gergs@rub.de}
\author{Thomas Mussenbrock}
\affiliation{Chair of Applied Electrodynamics and Plasma Technology, Department of Electrical Engineering and Information Science, Ruhr University Bochum, 44801 Bochum, Germany}
\author{Jan Trieschmann}
\affiliation{Theoretical Electrical Engineering, Department of Electrical and Information Engineering, Kiel University, Kaiserstraße 2, 24143 Kiel, Germany}
\email[]{jt@tf.uni-kiel.de}

\begin{abstract}
Molecular dynamics simulations are often used to study sputtering and thin film growth. Compressive stresses in these thin films are generally assumed to be caused by a combination of forward sputtered (peened) built-in particles and entrapped working gas atoms. While the former are assumed to hold a predominant role, the effect of the latter on the interaction dynamics as well as thin film properties are scarcely clarified (concurrent or causative). The inherent overlay of the ion bombardment induced processes render an isolation of their contribution impracticable. In this work, this issue is addressed by comparing the results of two case studies on the sputter deposition of Al thin films in Ar working gas. In the first run Ar atoms are fully retained. In the second run they are artificially neglected, as implanted Ar atoms are assumed to outgas anyhow and not alter the ongoing dynamics significantly. Both case studies have in common that the consecutive impingement of 100 particles (i.e., Ar$^+$ ions, Al atoms) onto Al(001) surfaces for ion energies in the range of 3 eV to 300 eV as well as Al/Ar$^+$ flux ratios from 0 to 1 are considered. The surface interactions are simulated by means of hybrid reactive molecular dynamics/force-biased Monte Carlo simulations and characterized in terms of mass density, Ar concentration, biaxial stress, shear stress, ring statistical connectivity profile, Ar gas porosity, Al vacancy density, and root-mean-squared roughness. Ultimately, implanted Ar atoms are found to form subnanometer sized eventually outgassing clusters for ion energies exceeding 100 eV. They fundamentally govern a variety of surface processes (e.g., forward sputtering/peening) and surface properties (e.g., compressive stresses) in the considered operating regime.
\end{abstract}

\maketitle

\newpage


\section{Introduction}
\label{sec:introduction}

Physical vapor deposition (PVD) refers to a group of methods that are based either on the evaporation or sputtering of atoms from a target material, their transport through the gas phase and the subsequent growth at the substrate. Sputtering describes the ejection of surface atoms as a result of a collision cascade that is spawned by an energetic ion bombardment. The projectiles are often mainly ions created from the inert background gas (e.g., argon). The ions are accelerated towards the walls when leaving the plasma bulk by the electric field due to the electric potential drop in the plasma sheaths. This manifests in minor or substantial kinetic energies at the substrate or target, depending on this electric potential difference. In addition to the removal of surface atoms, the energetic ions eventually cause densifications of the just grown thin films, as well as forward sputtered (peened) point defects \citep{abadias2018stress,d1989note,windischmann1992intrinsic,sigmund1981sputtering}. These defects are assumed to be the predominant reason for compressive stresses within sputter deposited thin films. The concurrent incorporation of heavy inert gas atoms causes distortions of the surrounding lattice structure and provides recoil centers that result in higher sputter yields. However, detailed investigations of their holistic role during sputtering or growth processes are unavailable to the best of our knowledge \citep{abadias2018stress,windischmann1992intrinsic}.. 

A simple scenario is used in this work to facilitate the required isolation of effects that are related to the entrapment of the working gas, that is Ar in sputter deposited Al thin films. It is particularly relevant, as Al is also utilized as one elemental or composite target material for the sputter deposition of hard ceramics, e.g., TiAlON and VAlON \citep{gibson2018quantum}.

Ion implantation studies revealed that incorporated Ar atoms tend to form bubbles within the Al matrix due to their insolubility \citep{vom1984pressure,tyagi1986solid,donnelly1986lattice,godet2018depth,dhaka2010plasmon,dhaka2008xe}. However, substantially higher ion energies compared to the typical case of PVD are used and this phenomena is argued to be ion energy and dose dependent \citep{biswas2004argon}. Hence, its occurrence may be less prominent during sputter deposition. Growth and sputtering processes are often investigated by means of classical or reactive molecular dynamics (RMD) simulations \citep{neyts2017molecular,graves2009molecular}. The former has been used by Valkealahti and Nieminen to investigate the Ar bombardment of Al surfaces with kinetic energies of 200 eV and 400 eV. Subsequent to each individual collision cascade, the temperature of the total surface slab has been adjusted by velocity rescaling over the course of 200 timesteps. The Morse, Lennard-Jones (LJ) and Ziegler-Biersack-Littmark (ZBL) potentials have been used for Al-Al, Ar-Ar and Al-Ar pair interactions, respectively. \citep{valkealahti1986molecular}

In this work, the scenario of Ar ion impingement and incorporation on Al(100) is studied for two cases as outlined below. The applied methods are described in Section~\ref{sec:methods}, wherein specifically the interaction potentials are summarized in Section~\ref{ssec:interaction_potentials}. The considered case studies (i.e., with and without implanted Ar atoms) as well as the routine for modelling the consecutive impingement of particles is detailed in Section~\ref{ssec:surface_interaction}. Section~\ref{ssec:evaluation_sproperties} covers the corresponding evaluation procedure. The results of both case studies are then compared and discussed in Section~\ref{sec:results}. The effect of implanted Ar atoms on the surface state and ongoing dynamics is assessed. Finally, in Section~\ref{sec:conclusion} conclusions are presented.

\section{Methods}
\label{sec:methods}

The open source molecular dynamics code LAMMPS is used to simulate the surface interactions considered in this work \citep{plimpton1995fast,thompson2022lammps}. The therefor used and combined interaction potentials (i.e., COMB3, LJ, ZBL) are laid out first. Second, the case studies (i.e., with and without implanted Ar atoms) are defined. The corresponding surface interactions are subdivided into individual model steps and highlighted consecutively. Finally, the methodology for the evaluation of the surface state (i.e., density, stoichoimetry, stress, defect structure) and some additional information (i.e., Ar gas porosity, Al vacancy density, root-mean-squared roughness) are presented.

\subsection{Interaction potentials}
\label{ssec:interaction_potentials}

Aluminum complexes are modeled by means of the Charge-Optimized Many-Body (COMB3) potential, which consists of electrostatic and short-range interactions, as well as contributions depicted by Legendre polynomial terms \citep{choudhary2014charge,liang2013classical}. The charge of any atom is assumed to be described by an effective point core charge $Z$ in combination with a rigid 1s Slater type orbital for the remaining charge $q-Z$, $q$ being the net charge of the atom. The short-range interaction is defined by the sum of a pairwise repulsion and scaled attraction term. This scale is the mean bond order of both atoms, being functions of their individual local atomic environment. The contributions due to the Legendre polynomials are meant to favor certain bond angles energetically and, hence, ease for instance the distinction of face-centred cubic (fcc) and hexagonal close-packed (hcp) phases. A more thorough description of the COMB3 formalism can be found elsewhere \citep{liang2013classical}. 

A combination of the COMB3 with the ZBL potential is introduced to account for high-energy collisions during the ion bombardment \citep{ziegler1985stopping}. All two or many body potential functions are modified by tapering with $1-f_\mathrm{C}$ to have them gradually decrease to zero for pair distances from 2 \r A to 1 \r A, $f_\mathrm{C}$ being the Tersoff cutoff function. Correspondingly, the ZBL function is also scaled with $f_\mathrm{C}$ to continuously increase its impact and have it eventually take over once the distance between two atoms becomes smaller than 1 \r A. However, the routine for the bond order computation is chosen to not be affected by this modification. This means that two Al atoms with a pair distance of less than 1 \r A still are aware of each other bond-order-wise. 

The LJ potential (with a cutoff radius of 8.5 \r A) and the ZBL potential are tapered in a similar manner from 2.25 \r A to 3.05 \r A, utilizing again the Tersoff cutoff function as tapering function \citep{ziegler1985stopping}. The depths of the LJ potential well $\epsilon$ are 0.02026 and 0.0104 for Al-Ar and Ar-Ar atom pairs, respectively. The zero-crossing distances $\sigma$ are 3.01 \r A and 3.40 \r A for Al-Ar and Ar-Ar interactions, respectively \citep{sha2018molecular,valkealahti1986molecular}.

The implementations are verified by comparing the integrated force with the corresponding potential energy as well as the tapered with the non-tapered potentials for a variety of cases. 

\subsection{Surface interaction}
\label{ssec:surface_interaction}

In the following, the workflow, the methods, and the inherent approximations are presented that are used during the surface interaction simulations. Initially general considerations are discussed. Thereafter, an iterative scheme is proposed to loop through the process of individual particle-surface interactions.

The incoming particle fluxes as well as the simulation setups are defined as follows. The case study consists of 5 different $\Gamma_\mathrm{Al}^\mathrm{in}/\Gamma_\mathrm{Ar^+}^\mathrm{in}$ flux ratios (i.e., 0.0, 0.25, 0.5, 0.75 and 1.0), as well as 20 ion energies in the range of 3 eV to 300 eV. Thermal Ar neutrals (Al ions) are assumed to have a negligible effect on the surface dynamics due to their low momentum (rare occurrence), respectively. Both species are neglected in this work. Each case consists of the consecutive impingement of 100 particles, so that in total 10000 surface interactions with 100 different energy/flux ratio combinations are studied.

The Ar$^+$ ion energies of the different cases are distributed equidistantly following a square root energy axis (resembling a linear axis in velocity/momentum and related processes). Hence, the considered list of ion energies starts with [3 eV, 6.52 eV, 11.38 eV, \dots] and ends at 300 eV. Ar ions are assumed to hit the Al(001) surface perpendicularly with both lateral velocity components equal to zero. The surface dimensions are chosen as a function of the impinging ion energies, that is $5\times5\times6,~6\times6\times7,~7\times7\times8$, and $8\times8\times9$ unit cells for ion energies in the range of [3 eV, 50 eV], (50 eV, 150 eV], (150 eV, 250 eV] and (250 eV, 300 eV], respectively. 

The required lattice constant at the given temperature (i.e., 300 K) is evaluated by performing a simulation of an Al bulk system consisting of a $8\times8\times8$ sized supercell over the course of 500 ps with a timestep of 0.2 fs. The open-source molecular dynamics code LAMMPS \citep{plimpton1995fast,thompson2022lammps} and the therein implemented Nose-Hoover style non-Hamiltonian equations of motion are utilized to determine the lattice constant to be 4.065 \r A by sampling from the isothermal-isobaric (NPT) ensemble during the production run (i.e., within the last 200 ps) \citep{plimpton1995fast,thompson2022lammps,dullweber1997symplectic,martyna1994constant,shinoda2004rapid,tuckerman2006liouville,parrinello1981polymorphic}. The damping constants for the thermostat and barostat are chosen to be 0.1 ps and 1.0 ps, respectively.

The Al(001) surface slab is created by cleaving the bulk system in [001] direction and increasing the simulation box height by 21 \r A, out of which 11 \r A are equal to the COMB3 cutoff radius and 10 \r A serve as a buffer for recognizing sputtered particles. Two bottom layers of the slab are excluded from the time integration and, hence, are kept immobile. The remaining, active Al atoms are equilibrated at the targeted temperature (i.e., 300 K) for 50 ps by applying a Langevin thermostat with damping coefficient of 0.1 ps \citep{schneider1978molecular,dunweg1991brownian}. Ar atoms are excluded from the thermostats. The Langevin thermostat is chosen here and in the following in favor of, for instance, the Nose-Hoover and Berendsen thermostat. This is reasoned by the circumstance that the former is sampling the system's internal state and, hence, equilibrates less well for a varying number of degrees of freedom, while the latter provides only a bias for matching the temperature but not necessarily for fulfilling the equipartition theorem. In the frame of the Langevin thermostat, however, a viscous background medium is implied, which drags kinetic energy from each atom individually and heats them up by having solvent atoms randomly bump into them. It has been argued that this and the required parameters can be interpreted and reasoned by means of electron phonon coupling \citep{hou2000deposition}. A varying number of degrees of freedom provides no issues and the thermostat pushes the system inherently towards the fulfillment of the equipartition theorem, at the cost of possibly altered vibrational motion. 

\begin{figure}[t]
\includegraphics[width=8cm]{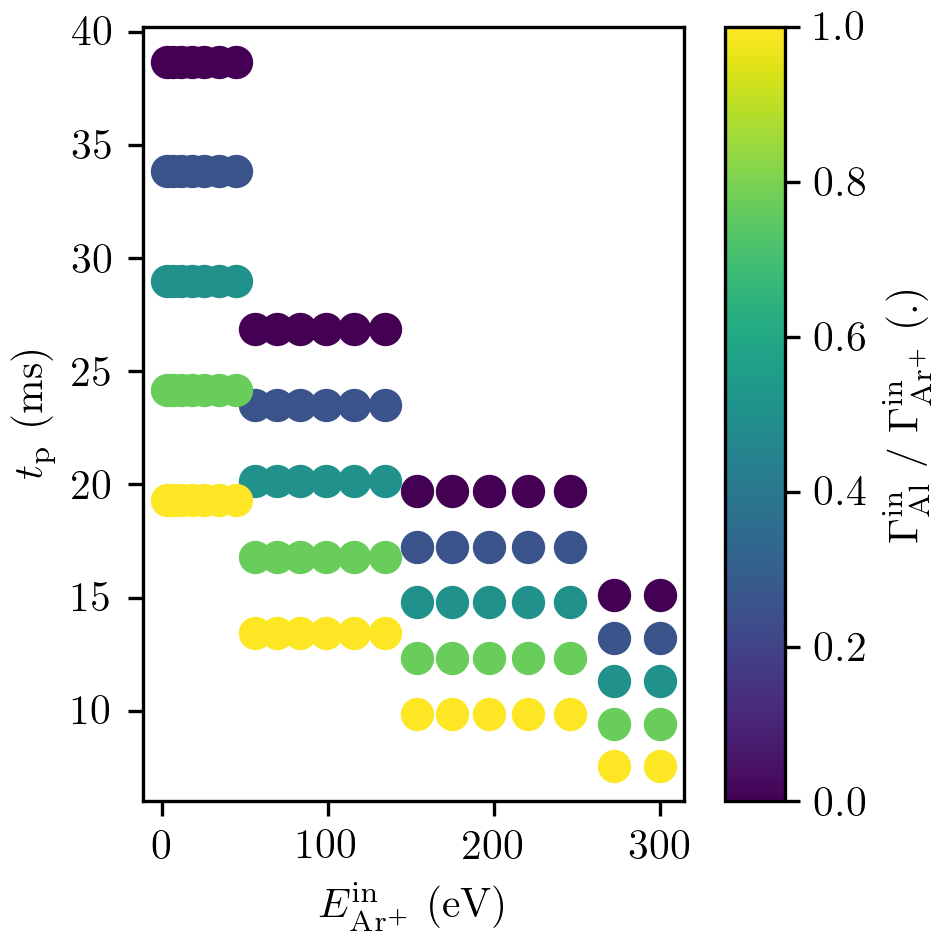}
\caption{The estimated process time $t_\mathrm{p}$ for a typical dcMS Al thin film deposition ($I$=348 mA, $P$= 100 W, $A_\mathrm{t}$= 63.6 cm$^2$ detailed elsewhere \citep{trieschmann2018combined}) is displayed as a function of the Ar ion energy $E_\mathrm{Ar^+}^\mathrm{in}$ as well as the ratio of the Al/Ar$^+$ fluxes towards the surface $\Gamma_\mathrm{Al}^\mathrm{in}/\Gamma_\mathrm{Ar^+}^\mathrm{in}$.}
\label{fig:process_time}
\end{figure}

The process time for the different energy/flux ratio combinations may be estimated by dividing the number of bombarding Ar ions by the Ar ion impinging rate, $t_\text{p} = N (1-0.5\Gamma_\mathrm{Al}^\mathrm{in}/\Gamma_\mathrm{Ar^+}^\mathrm{in}) / (\Gamma_\mathrm{Ar^+}^\mathrm{in}A_\mathrm{RMD})$. Therein, $N=100$ is the total number of projectiles and $A_\mathrm{RMD}$ is the considered RMD surface area. A typical Ar ion flux onto the target for the example of direct current magnetron sputtering (dcMS) and Al thin film deposition \cite{trieschmann2018combined} can be estimated as a function of the current $I$=348 mA, the target area $A_\mathrm{t}$= 63.6 cm$^2$ and the ion-induced secondary electron emission coefficient $\gamma_\mathrm{SE}$=0.091 following $\Gamma_\mathrm{Ar^+}^\mathrm{in} = w_\mathrm{rt} I/((1+\gamma_\mathrm{SE})eA_\mathrm{t})=6.26 \cdot 10^{20}~\mathrm{m}^{-2}\mathrm{s}^{-1}$ \citep{trieschmann2018combined,DEPLA20064329}. Therein $w_\mathrm{rt} \approx 2$ is a scaling factor taking into account the inhomogeneous ion flux distribution over the racetrack. The estimated process time is depicted in Figure~\ref{fig:process_time}.

\subsubsection{Atom generation}
\label{sssec:atom_generation}

The dimensions of the simulation box are initially set up as a function of energy as specified above. The simulation domain is subsequently extended in vertical direction to include 21 \r A above the uppermost Al atom. The impinging particle (i.e., Al or Ar$^+$) is positioned 11 \r A above this Al atom (10 \r A below the upper simulation box bound). However, if the atom would take longer than 1 ps to reach the surface (i.e, its vertical velocity component $v_z$ is smaller than -11 \r A/ps) the position is lowered to 3.7 \r A+$v_z \cdot 1$ ps above the uppermost Al atom. This means that the atom is put within the surface atoms' interaction radii. However, the lower bound of 3.7 \r A ensures that at least no Al atom instantaneously contributes any kind of short range interaction, whose outer cutoff radius is 3.7 \r A for Al complexes. The lateral coordinates are sampled from uniform distributions over the unit interval U(0,1) and scaled with the particular simulation box dimensions. 

The absolute velocity $v$ of individual (sputtered) Al neutrals are sampled from  
\begin{equation}
\label{eq:velocity_sampling}\\
v = \frac{\sqrt{-\mathrm{ln}(\mathrm{U(0,1)})}}{\beta}
\end{equation}
with $\beta = \sqrt{\frac{m}{2k_\mathrm{b}T_\mathrm{sp}}}$ being the inverse of the most probable molecular speed (flux towards the surface), $m$ is the particular atom mass, $k_\mathrm{b}$ is the Boltzmann constant and $k_\mathrm{b}T_\mathrm{sp}$ refers to the kinetic energy of sputtered and subsequently thermalized Al atoms, which is estimated to be approximately 7 eV by averaging the Sigmund--Thompson distribution \citep{bird1994molecular,thompson1968ii,sigmund1969theory,sigmund1969theory2}. The azimuthal angle $\phi$ is sampled from a scaled U(0,1) distribution, $\phi = 2\pi$U(0,1). The proportion of the $z$ component (in surface normal direction) is determined by $\cos(\theta)=\mathrm{U}(0,1)$ and $v_z=\cos(\theta) v$. The polar component $\sin(\theta)$ and the lateral components $v_x$ as well as $v_y$ are consequently computed by $v_x=\cos(\phi)\sin(\theta)v$ and $v_y=\sin(\phi)\sin(\theta)v$ with $\sin(\theta)=\sqrt{1-\cos(\theta)^2}$.

The species of each individual atom is randomly determined for the given flux ratio by applying Monte Carlo accept-reject sampling (rejection sampling).

\subsubsection{Impact and relaxation}
\label{sssec:impact_and_relaxation}

During the phase of particle approach and impact, the timestep size is adjusted every 100 steps with an upper bound of 0.2 fs to ensure that the maximum displacement of any atom does not exceed 0.01 \r A per timestep. To avoid an alteration of the sputtering dynamics due to the applied Langevin thermostat, atoms within a sphere with radius $r_\mathrm{th}$ around the impinging atom are excluded from the thermostat. Once excluded, they remain so during this phase. The radius $r_\mathrm{th}$ is 8.0 \r A, 10 \r A, 12 \r A and 14 \r A for ion energies in the range of [3 eV, 50 eV], (50 eV, 150 eV], (150 eV, 250 eV] and (250 eV, 300 eV], respectively. The sphere and list of bypassed atoms is updated every 100 timesteps, which allows for a maximum displacement of any atom of 1 \r A. 

Sputtered, desorbed, or reflected atoms that approach the upper bound of the simulation box by less than 10 \r A are excluded from the time integration and, hence, are kept immobile to evaluate them later. The simulation time for the impact and the corresponding relaxation of the eventually spawned collision cascade is chosen to be a function of the individual kinetic energy of each atom: 1.5 ps + 0.005$E_\mathrm{kin}$ ps/eV.

For the second case study, any implanted Ar atom is removed from the system, assuming that the majority of them outgas anyhow or do not alter the surface state and interactions significantly. Its influence will be detailed later.

\subsubsection{Relaxation and diffusion}
\label{sssec:relaxation_and_diffusion}

This phase is meant to relax artificial, high pressure that is a consequence of introducing a new particle every few picoseconds to the system and to account for diffusion processes by applying the time-stamped force-bias Monte Carlo (tfMC) method \citep{neyts2014combining,bal2014time,mees2012uniform}. The force acting on each individual atom is therein used to sample the atoms' displacements without integrating Newton's equations of motion. As a result, longer simulation times per step may be reached by utilizing tfMC in favor of RMD. The system wide parameter $\Delta$, which represents the maximum displacement of the lightest atom (i.e., Al) during a single tfMC step is chosen to be 0.2 \r A (approximately 10 \% of the nearest neighbor distance in case of an Al interstitial). The corresponding value for Ar atoms is scaled with the square root of the mass ratio, $\sqrt{\frac{m_\mathrm{Al}}{m_\mathrm{Ar}}}$. The simulation is run for $10^4$ steps at 300 K.

\subsubsection{Evaluation of desorbed, sputtered, and reflected particles}
\label{sssec:diagnosing}

To differentiate between surface and gas phase atoms, an instantaneous cluster analysis with a cutoff radius of 3.3 \r A is performed. The group of surface atoms is updated and the type as well as velocities of the gas phase atoms are stored prior to their removal from the system.

\subsubsection{Phase space relaxation}
\label{sssec:phase_space_relaxation}

The tfMC simulation implies distortions in the phase space, which are larger than those that typically occur during RMD steps. Hence, the atoms have to be relaxed to obtain a suited phase space representation before another atom impinges the surface. An artificial accumulation of defects as well as an enhancement of sputtered particles may otherwise be the consequence. This is dealt with by coupling the whole mobile system to the Langevin thermostat again and performing a simulation for 0.5 ps. The resulting system state is then written to a binary restart file to evaluate the surface properties later, after the surface interactions of 50 and 100 particles have been simulated.

\subsection{Evaluation of surface properties}
\label{ssec:evaluation_sproperties}

The mobile Al atoms of the reloaded surface slab are coupled to a Nose-Hoover thermostat with a damping constant of 0.1 ps and are equilibrated for 100 ps, out of which the last 50 ps are used to evaluate the herein considered surface properties. The short simulation time of the equilibration phase and the  production run are a compromise to reduce the overall computational burden. Hence, the obtained properties may be regarded as estimates. 

The surface state is assumed to be described by the stoichoimetry, density, stress, and defect structure. The first quantity is described by the particular Ar concentration $x_\mathrm{Ar}$. The mass density $\rho$ is computed by dividing the net mass of the system by the corresponding volume. The latter is obtained from the point-like particles by utilizing the alpha-shape algorithm with a cutoff radius of 3.3 \r A, which is implemented in the OVITO python module \citep{stukowski2009visualization,stukowski2014computational}. This volume is also used to extract the biaxial stress $\frac{\sigma_{xx}+\sigma_{yy}}{2}$ and the shear stress $\tau_{xy}$ from the volume times stress tensor product, which has been computed with LAMMPS during the production run \citep{plimpton1995fast,thompson2022lammps,thompson2009general}. The defect structure is described by means of ring statistics. Rings are closed paths that consist of nodes (atoms) and bonds. Two Al atoms that are less than 3.0 \r A apart from each other are considered as neighbors and, hence, are interconnected with a bond. This does not imply any ionic or covalent characteristic and is solely used for the network analysis. The strong rings criterion is applied and means that no ring can be described by the combination of multiple smaller rings \citep{yuan2002efficient,goetzke1991properties}. This allows for a continuous description from zero dimensional point defects (e.g., vacancies, interstitials, anti-sites) up to three dimensional amorphous defect clusters. The ring networks are described by the connectivity profile of rings with $n~\in~[3,6]$ nodes. It consists of the number of rings per cell normalized by the number of nodes in the network $R_{\mathrm{C},n}$, the proportion of nodes at the origin of at least one ring with $n$ nodes $P_{\mathrm{N},n}$, the proportion of nodes for which the $n$ sized rings represent the shortest closed paths $P_{\mathrm{min},n}$ and longest closed paths $P_{\mathrm{max},n}$ \citep{drabold2005models,cobb1996ab,zhang2000structural}. The connectivity profile is computed with the R.I.N.G.S. code \citep{le2010ring}.

In addition, the Ar gas porosity $\phi_\mathrm{Ar}$, the Al vacancy density $N_\mathrm{V_{Al}}$,  and the root-mean squared roughness $R$ are determined. The first quantity is described by the particular Ar fraction of the surface slab's volume, whose computation is based on the alpha-shape algorithm introduced in the preceding paragraph \citep{stukowski2009visualization,stukowski2014computational}. The same methodology is applied to each species separately. The volume which is occupied by Ar atoms is computed as the average of 
the Ar atoms' volume and the total surface slab's volume minus the Al atoms' volume. The Al vacancy density is the number of vacant fcc lattice sites (amorphous regions are neglected) normalized by the beforehand mentioned slab volume. Vacant Al sites that are occupied by Ar atoms are not considered as Al vacancies. However, their occurrence is quantified by means of the connectivity profile (defect structure characterization). The root-mean squared roughness is computed as the square root of the ensemble averaged variance of the surface atoms' surface normal coordinate \citep{lou2004feedback}. This group of atoms is also identified with the alpha-shape algorithm.

\section{Results}
\label{sec:results}

The case study outlined in section \ref{ssec:surface_interaction} is performed twice: (i) maintaining the implanted Ar atoms as is and (ii) removing implanted Ar atoms from the system in the second run and assuming that those atoms either outgas anyhow or have a negligible effect on the subsequent dynamics. The validity of these assumptions as well as the role of Ar ions in general for Al-based sputter deposited thin films is meant to be addressed in this section by comparing the results of both cases for the impingement of the 1st and 2nd set of 50 particles each.

\begin{figure}[t]
\includegraphics[width=8cm]{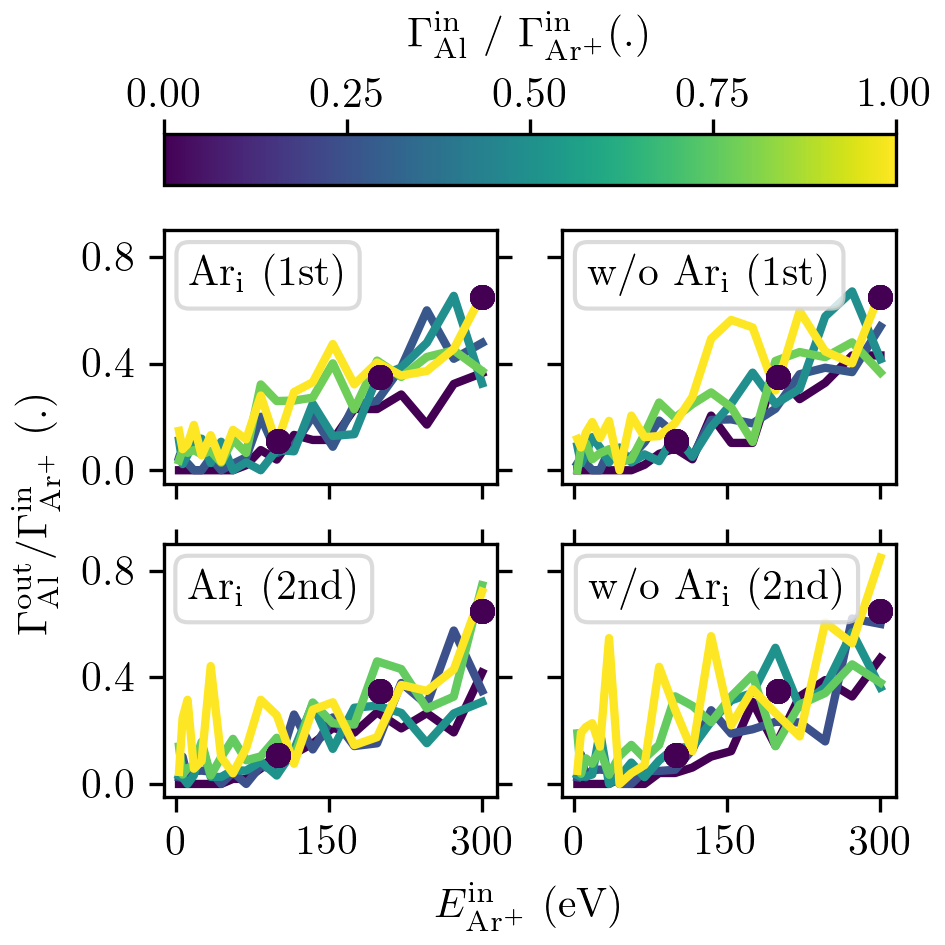}
\caption{The Al flux from the Al(001) surface $\Gamma_\mathrm{Ar}^\mathrm{out}$ per Ar$^+$ ion flux towards the surface $\Gamma_\mathrm{Ar^+}^\mathrm{in}$ is displayed as a function of the Ar ion energy $E_\mathrm{Ar^+}^\mathrm{in}$ and Al/Ar flux ratio $\Gamma_\mathrm{Al}^\mathrm{in}/\Gamma_\mathrm{Ar^+}^\mathrm{in}$ for the 1st and 2nd set of impinging particles as well as with and without implanted Ar atoms. Circles resemble experimental references for the sputter yield ($\Gamma_\mathrm{Al}^\mathrm{in}=0$) \citep{laegreid1961sputtering}.}
\label{fig:yield}
\end{figure}

The flux of Al atoms that have been sputtered or reflected $\Gamma_\mathrm{Al}^\mathrm{out}$ is normalized to the Ar$^+$ ion flux bombarding the surface $\Gamma_\mathrm{Ar^+}^\mathrm{in}$ for any herein considered ion energy and $\Gamma_\mathrm{Al}^\mathrm{in}/\Gamma_\mathrm{Ar^+}^\mathrm{in}$ flux ratio. The result is shown in Figure~\ref{fig:yield} for the scenarios with and without implanted Ar atoms. Both cases agree with each other and also with experimental findings \citep{laegreid1961sputtering}. For the case including Ar incorporation, the Ar atoms function as additional recoil centers, but also keep the surrounding Al atoms at a distance of approximately 3.0 \r A, which eventually distorts the fcc lattice structure locally.

\begin{figure}[t]
\includegraphics[width=8cm]{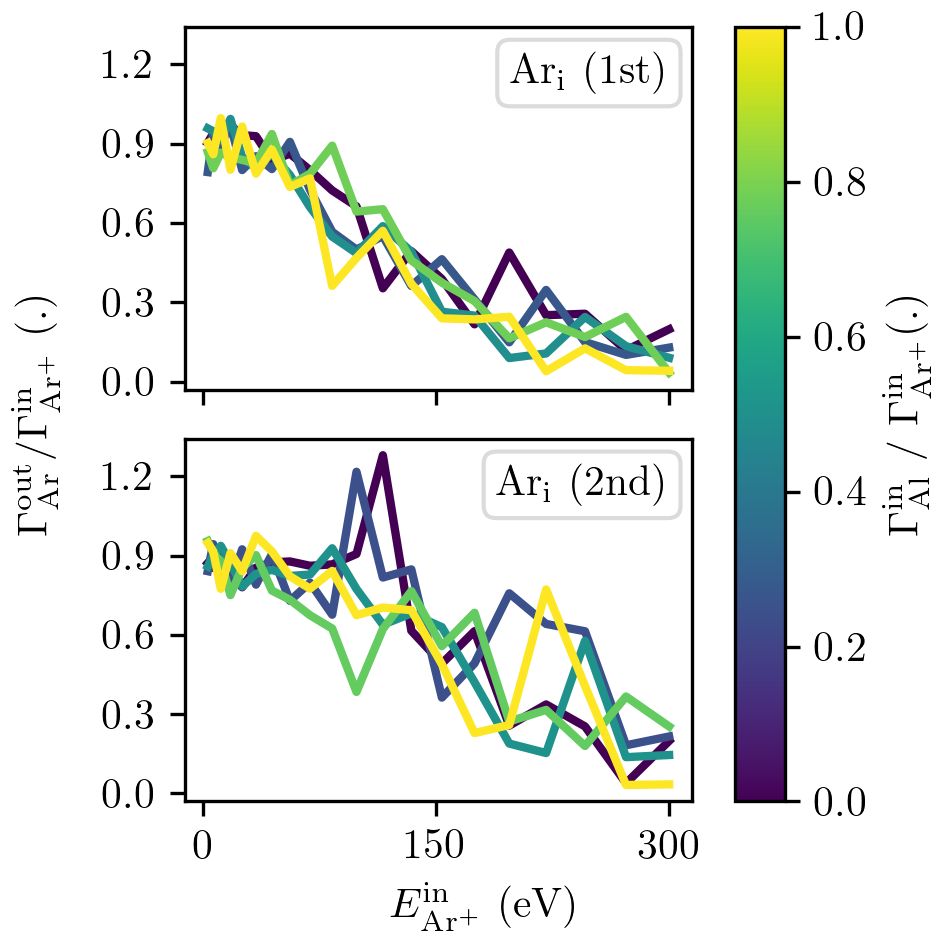}
\caption{The Ar flux from the Al(001) surface $\Gamma_\mathrm{Ar}^\mathrm{out}$ per Ar$^+$ ion flux towards the surface $\Gamma_\mathrm{Ar^+}^\mathrm{in}$ is displayed as a function of the Ar ion energy $E_\mathrm{Ar^+}^\mathrm{in}$ and Al/Ar flux ratio $\Gamma_\mathrm{Al}^\mathrm{in}/\Gamma_\mathrm{Ar^+}^\mathrm{in}$ for the 1st and 2nd set of impinging particles. Implanted Ar atoms are taken into account.}
\label{fig:Ar_yield}
\end{figure}

\begin{figure}[t]
\includegraphics[width=8cm]{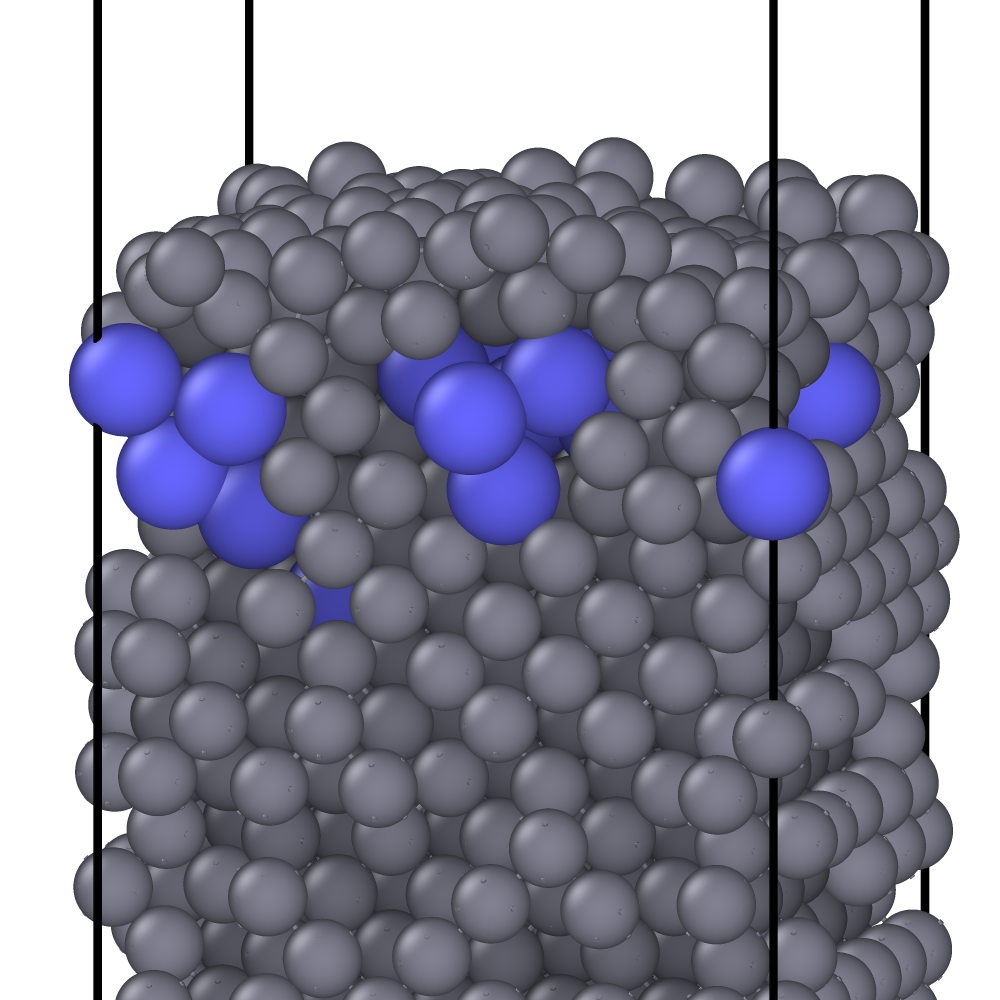}
\caption{Depiction of the atom structure prior to the outgassing of enclosed Ar clusters $\left(E_\mathrm{Ar^+}^\mathrm{in}=115.71~\mathrm{eV},~\Gamma_\mathrm{Al}^\mathrm{in}/\Gamma_\mathrm{Ar^+}^\mathrm{in}=0.0\right)$. Grey and purple spheres indicate Al and Ar atoms, respectively. The graphic is rendered with OVITO \citep{stukowski2009visualization}.}
\label{fig:ArBubbleRMD}
\end{figure}

For this case, the flux of Ar atoms that have been reflected or released from the surface $\Gamma_\mathrm{Ar}^\mathrm{out}$ is similarly normalized to the Ar$^+$ ion flux hitting the surface $\Gamma_\mathrm{Ar^+}^\mathrm{in}$ and shown in Figure~\ref{fig:Ar_yield}. Higher ion energies result in deeper and more probable Ar implantation. This likelihood is decreased during the 2nd set of impinging particles due to the gradual saturation of the surface slabs with enclosed Ar atoms. Notably, this process has still not reached a steady state, which would be indicated by a balance of incoming and outgoing Ar fluxes, $\Gamma_\mathrm{Ar}^\mathrm{out}\approx\Gamma_\mathrm{Ar^+}^\mathrm{in}$. The collision cascades caused by the Ar$^+$ ion bombardment allow for a continuous reordering that enables isolated and otherwise rather immobile Ar atoms to eventually form small clusters. Their size and depth is a function of the Ar concentration, distribution and, hence, of the Ar ion energy. At 98.73 eV and 115.71 eV these subnanometer sized clusters are covered by only one or two layer of Al atoms, which are gradually removed due to the ongoing ion bombardment until the entrapped Ar atoms outgas. This event causes the two spikes $\left(\Gamma_\mathrm{Ar}^\mathrm{out}/\Gamma_\mathrm{Ar^+}^\mathrm{in}>1\right)$ that are depicted in the 2nd set of Figure~\ref{fig:Ar_yield} for $\Gamma_\mathrm{Al}^\mathrm{in}/\Gamma_\mathrm{Ar^+}^\mathrm{in}$ flux ratios of 0 and 0.25. This reasoning is confirmed from the atom structure prior to the outgassing event as shown in Figure~\ref{fig:ArBubbleRMD}. For the remaining $\Gamma_\mathrm{Al}^\mathrm{in}/\Gamma_\mathrm{Ar^+}^\mathrm{in}$ flux ratios (i.e., 0.5, 0.75, 1.0) more Al atoms are deposited than sputtered from the surface per incoming Ar ion $\left(\Gamma_\mathrm{Al}^\mathrm{in}/\Gamma_\mathrm{Ar^+}^\mathrm{in}>\Gamma_\mathrm{Al}^\mathrm{out}/\Gamma_\mathrm{Ar^+}^\mathrm{in}\right)$ as presented in Figure~\ref{fig:yield}. Hence, the thin layers of Al atoms which cover the small Ar clusters become thicker over time. In case of higher ion energies a larger surface area (smaller dose) is considered and, furthermore, the Ar atoms are distributed deeper within the surface slab, so that it may take more time (impinging atoms) until the outgassing of the entrapped Ar clusters may be observed. The doses per set $\Gamma^\mathrm{in}t=50/A_\mathrm{RMD}$ for ion energies in the range of [3 eV, 50 eV], (50 eV, 150 eV], (150 eV, 250 eV] and (250 eV, 300 eV] are 12.10 atoms/\r A$^2$, 8.41 atoms/\r A$^2$, 6.18 atoms/\r A$^2$, 4.73 atoms/\r A$^2$, respectively.

\begin{figure}[t]
\subfloat[Ar concentration $x_\mathrm{Ar}$.\label{sfig:stoichiometry_wAr}]{%
    \includegraphics[width=8cm]{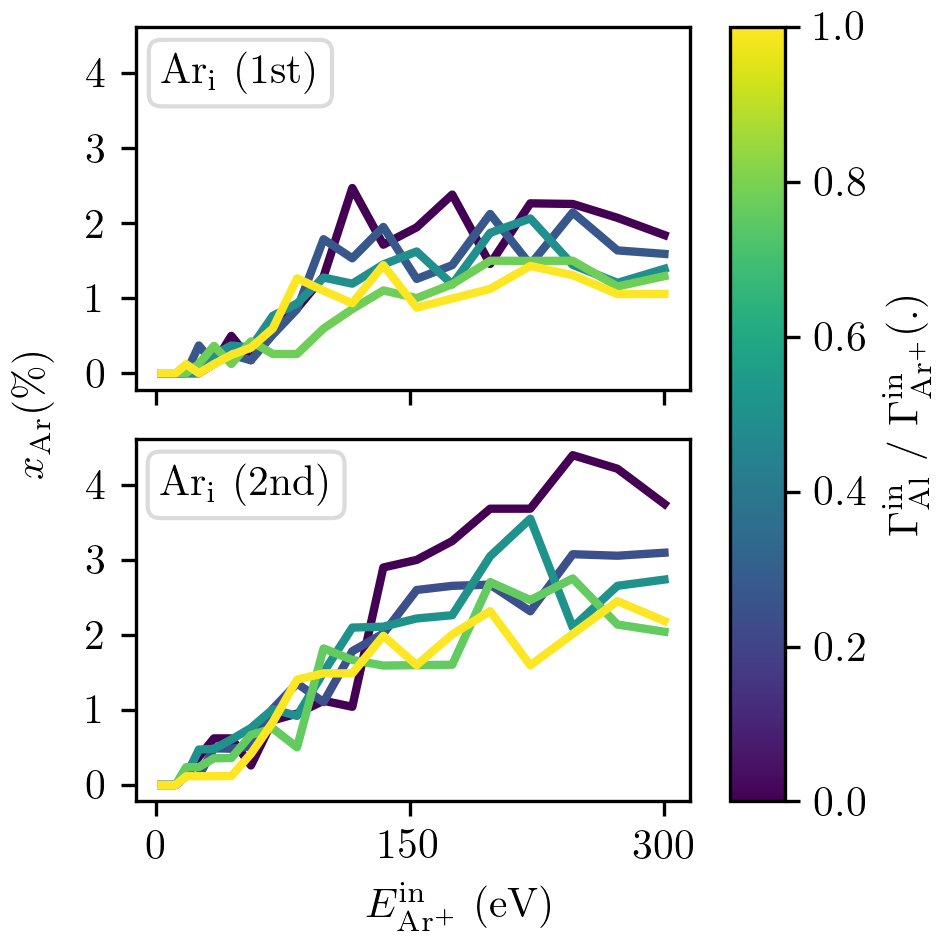}
}\hfill
\subfloat[Ar gas porosity $\phi_\mathrm{Ar}$. \label{sfig:ArGasPorosity_wAr}]{%
    \includegraphics[width=8cm]{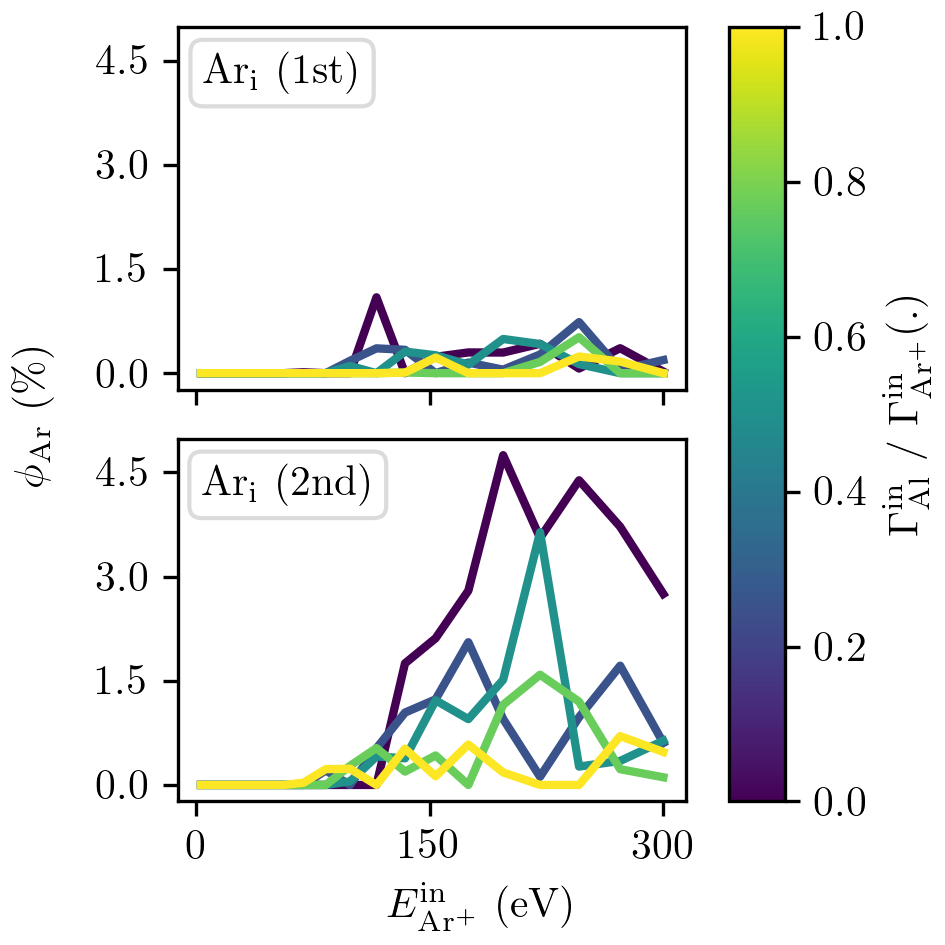}
}
\caption{The Ar concentration $x_\mathrm{Ar}$ and gas porosity $\phi_\mathrm{Ar}$ is displayed as a function of the Ar ion energy $E_\mathrm{Ar^+}^\mathrm{in}$ and Al/Ar flux ratio $\Gamma_\mathrm{Al}^\mathrm{in}/\Gamma_\mathrm{Ar^+}^\mathrm{in}$ for the 1st and 2nd set of impinging particles. Implanted Ar atoms are taken into account.}
\label{fig:ArConcAndGasPorosity}
\end{figure}

The corresponding Ar concentration $x_\mathrm{Ar}$ is shown in Figure~\ref{sfig:stoichiometry_wAr}. The change from the 1st to the 2nd set for cases with ion energies greater than approximately 100 eV indicate that those have not reached a steady state yet. This can also be noted by the dependency of the Ar concentration on the $\Gamma_\mathrm{Al}^\mathrm{in}/\Gamma_\mathrm{Ar^+}^\mathrm{in}$ flux ratio and, hence, Ar$^+$ ion dose (more bombarding Ar ions lead to a larger fraction of implanted Ar atoms). This conclusion is in agreement with the previous observation.

The Ar atoms are either isolated or partially clustered within the surface slab. The volume fraction which is occupied by Ar clusters is expressed by means of the gas porosity $\phi_\mathrm{Ar}$ and shown in Figure~\ref{sfig:ArGasPorosity_wAr}. At the end of the 1st set, the majority of Ar atoms are distributed apart from each other. When the Ar concentration is further increased during the 2nd set, for cases with ion energies greater than approximately 100 eV the formation of Ar clusters is observed. The occupied volume is expected to decrease the density of the film. The Ar gas porosity and concentration have a similar dependency on the $\Gamma_\mathrm{Al}^\mathrm{in}/\Gamma_\mathrm{Ar^+}^\mathrm{in}$ flux ratio. It may be inferred that a critical Ar concentration is required for the formation of subnanomenter sized Ar clusters beneath the surface layer. This concentration is found to be approximately 2 \%, which notably relates to the supercell dimensions in surface normal direction chosen here. 

\begin{figure}[t]
\subfloat[Al vacancy density $N_\mathrm{V_\mathrm{Al}}$.\label{sfig:AlVacancyDensity}]{%
    \includegraphics[width=8cm]{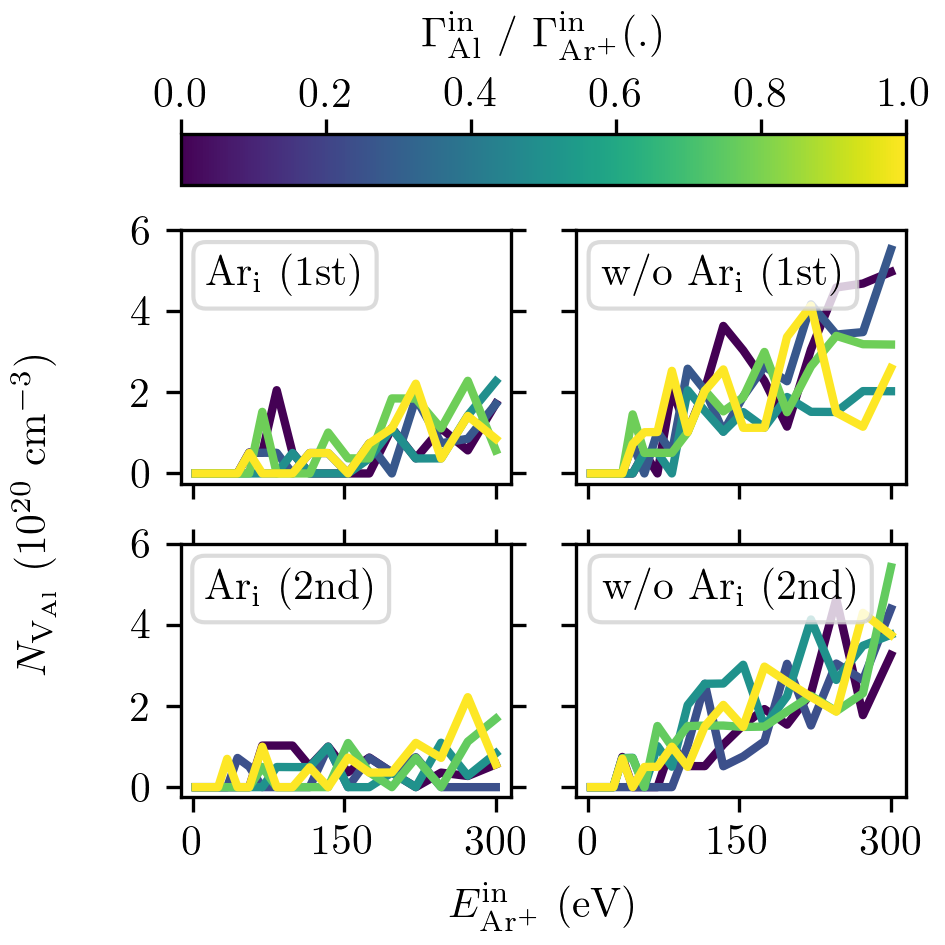}
}\hfill
\subfloat[Mass density $\rho$. \label{sfig:density}]{%
    \includegraphics[width=8cm]{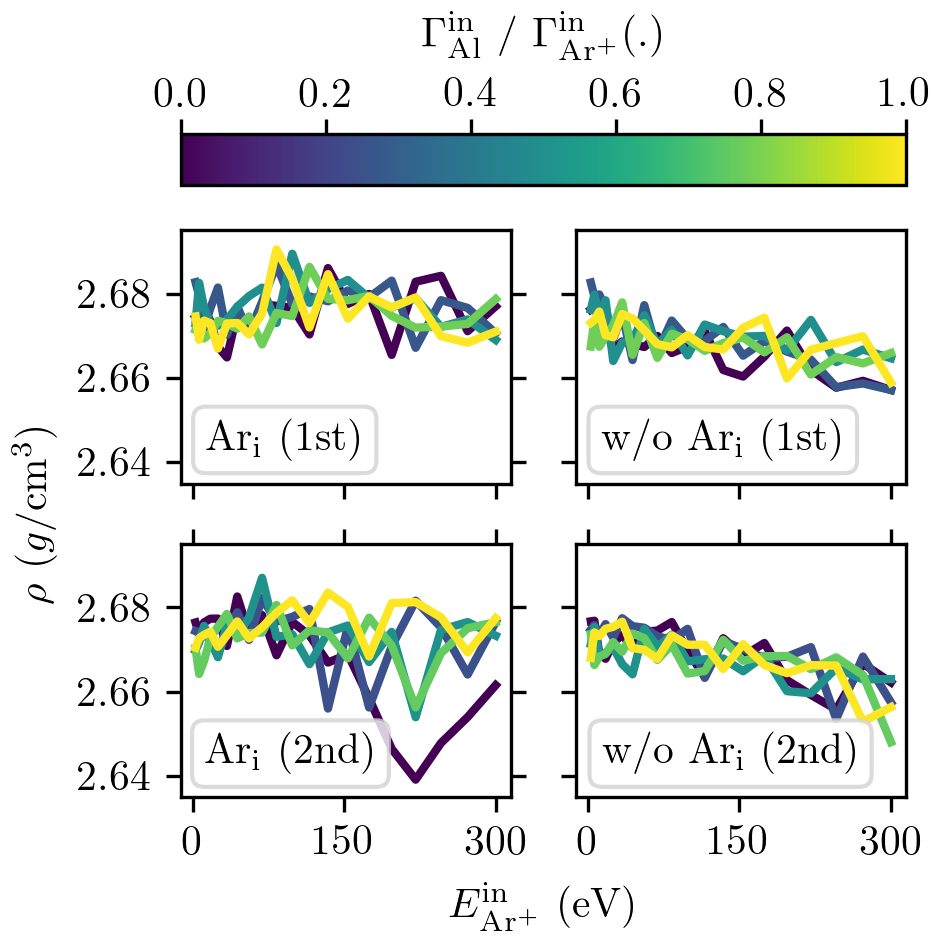}
}
\caption{The Al vacancy density $N_\mathrm{V_\mathrm{Al}}$ and mass density $\rho$ is displayed as a function of the Ar ion energy $E_\mathrm{Ar^+}^\mathrm{in}$ and Al/Ar flux ratio $\Gamma_\mathrm{Al}^\mathrm{in}/\Gamma_\mathrm{Ar^+}^\mathrm{in}$ for the 1st and 2nd set of impinging particles as well as with and without implanted Ar atoms.}
\label{fig:VacancyAndDensity}
\end{figure}

The bombarding of the surface with energetic ions leads to collision cascades which in addition to sputtering of surface atoms and implantation of Ar atoms cause a series of Al Frenkel defects. The majority of these vacancies and interstitials recombine, but some Al atoms are moved atop so that the remaining Al vacancies are the most dominant point defect structure. The corresponding porosity is however also diminished by recoil implantation and forward sputtering (peening) which densifies the surface slab. The Al vacancy density $N_\mathrm{V_\mathrm{Al}}$ is shown in Figure~\ref{sfig:AlVacancyDensity}. Higher ion energies lead to more Frenkel pairs and, hence, to more Al vacancies. This trend can be observed for both cases with and without Ar implantation, but is more prominent when incorporated Ar atoms are neglected. The latter function as recoil centers when colliding with Ar and Al atoms due to elastic and inelastic collisions, respectively. These, however, also cause a more even redistribution of the collision cascades' momentum and the formation of fewer Frenkel pairs. The atomic stress and strain due to the implanted Ar atom distribution further reduce the stability of such vacancy interstitial pairs and facilitate their recombination. 

The mass densities for cases with and without implanted Ar atoms are shown in Figure~\ref{sfig:density}. The density is decreased for higher ion energies in spite of the peening when enclosed Ar atoms are neglected due to the accumulation of more Al vacancies. In the opposite case retaining implanted Ar atoms, the smaller Al vacancy density in combination with the weight of the Ar atoms, their induced strain, and surface reconstruction leads to an approximately constant mass density for any herein considered ion energy and $\Gamma_\mathrm{Al}^\mathrm{in}/\Gamma_\mathrm{Ar^+}^\mathrm{in}$ flux ratio during the 1st set of impinging particles. During the 2nd set the system with implanted Ar become less dense for ion energies greater than 100 eV due to the clustering of otherwise isolated Ar atoms and the corresponding volumetric partitioning, i.e., Ar gas induced porosity (shown in Figure~\ref{sfig:ArGasPorosity_wAr}). 

\begin{figure}[t]
\subfloat[Biaxial stress $\frac{\sigma_{xx}+\sigma_{yy}}{2}$.\label{sfig:bistress}]{%
    \includegraphics[width=8cm]{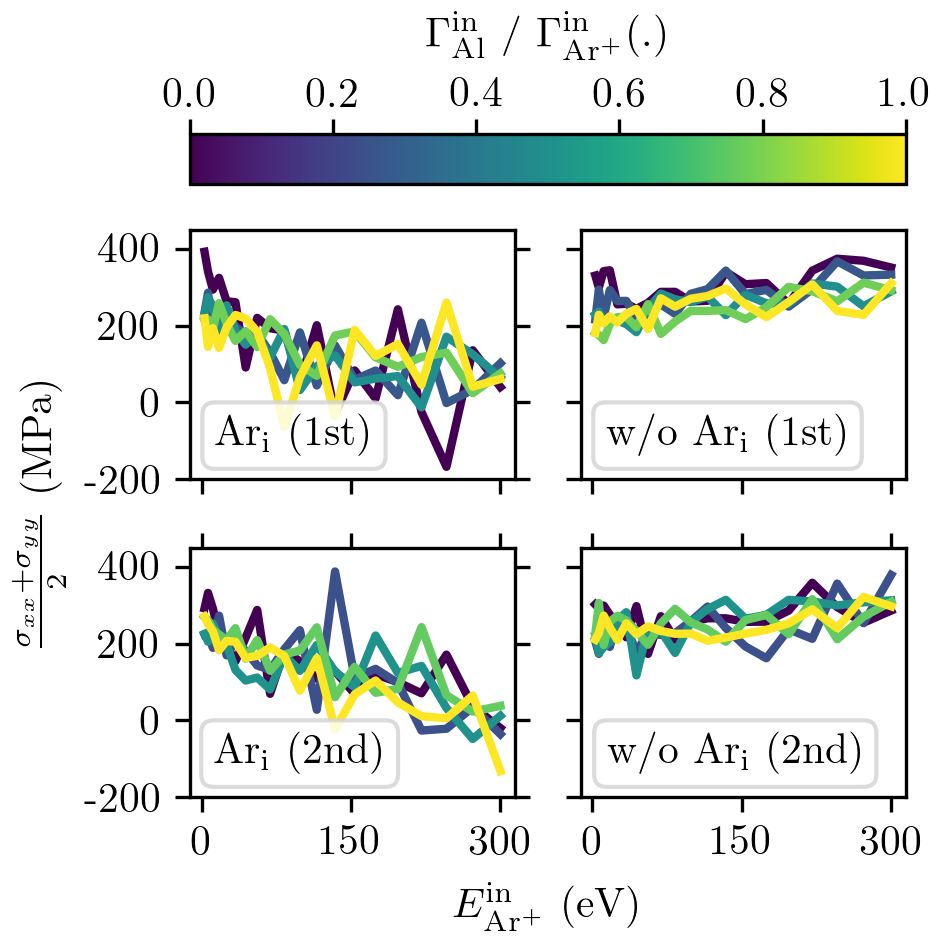}
}\hfill
\subfloat[Shear stress $\tau_{xy}$. \label{sfig:shstress}]{%
    \includegraphics[width=8cm]{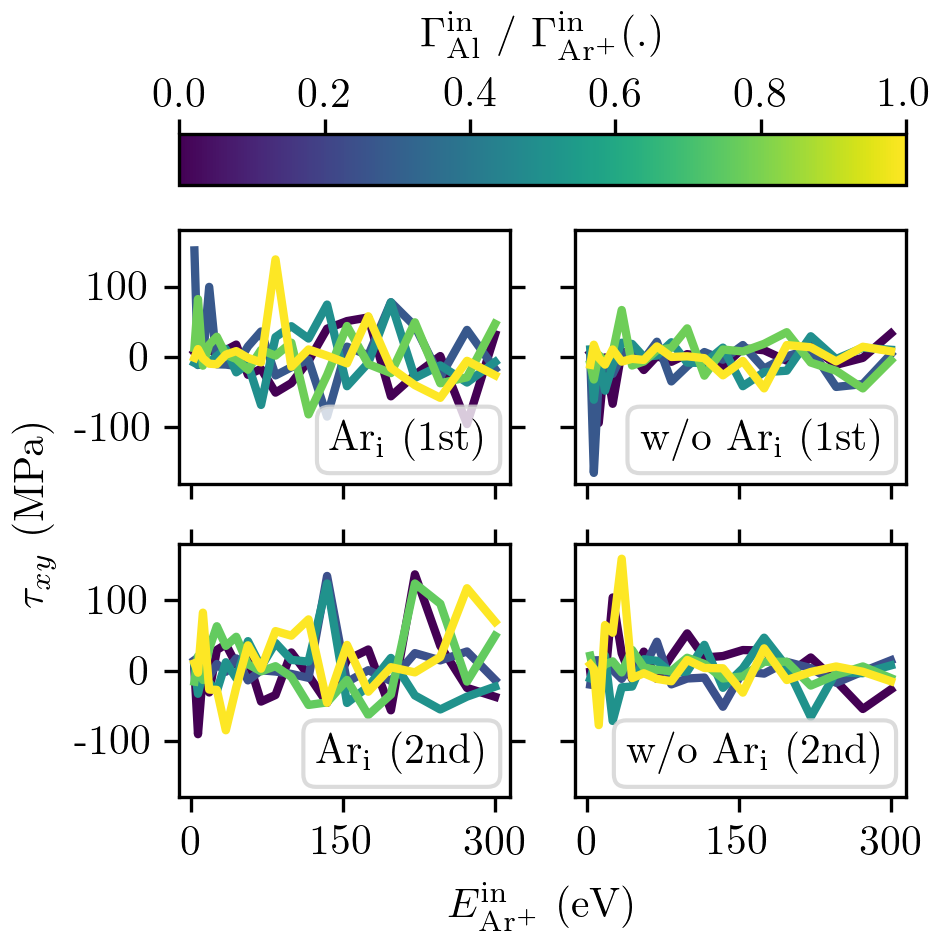}
}
\caption{The stress is displayed as a function of the Ar ion energy $E_\mathrm{Ar^+}^\mathrm{in}$ and Al/Ar flux ratio $\Gamma_\mathrm{Al}^\mathrm{in}/\Gamma_\mathrm{Ar^+}^\mathrm{in}$ for the 1st and 2nd set of impinging particles as well as with and without implanted Ar atoms.}
\label{fig:stress}
\end{figure}

The biaxial stress $\frac{\sigma_{xx}+\sigma_{yy}}{2}$ is shown in Figure~\ref{sfig:bistress} for cases with and without implanted Ar atoms. The biaxial stress is increased slightly for the latter scenario with higher ion energies, which is attributed to the increasing Al vacancy density (in spite of the peening process). Forward sputtered (peened) Al atoms (interstitials) predominantly recombine with vacant lattice sites. The surface slab experiences even more tensile stress for higher ion energies, even though the stress is expected to become compressive \citep{kim1998study,abadias2018stress,d1989note,windischmann1992intrinsic}. This trend is confirmed when enclosed Ar atoms are not neglected but taken into account. The cause is a combination of forward sputtered (peened) Al atoms (interstitials), porosity annihilation, and entrapped Ar atoms, whereas the last hinder the recombination of the first with the Al vacancies. Notably, surface interaction with a higher dose is required to obtain the steady state biaxial stress.

The shear stress $\tau_{xy}$ is shown in Figure~\ref{sfig:shstress} for cases with and without implanted Ar atoms. It is smaller than the particular biaxial stress in both scenarios. However, it does not reveal any particular trend, neither with/without implanted Ar atoms nor with ion energy or $\Gamma_\mathrm{Al}^\mathrm{in}/\Gamma_\mathrm{Ar^+}^\mathrm{in}$ flux ratio.

\begin{figure}[t]
\includegraphics[width=8cm]{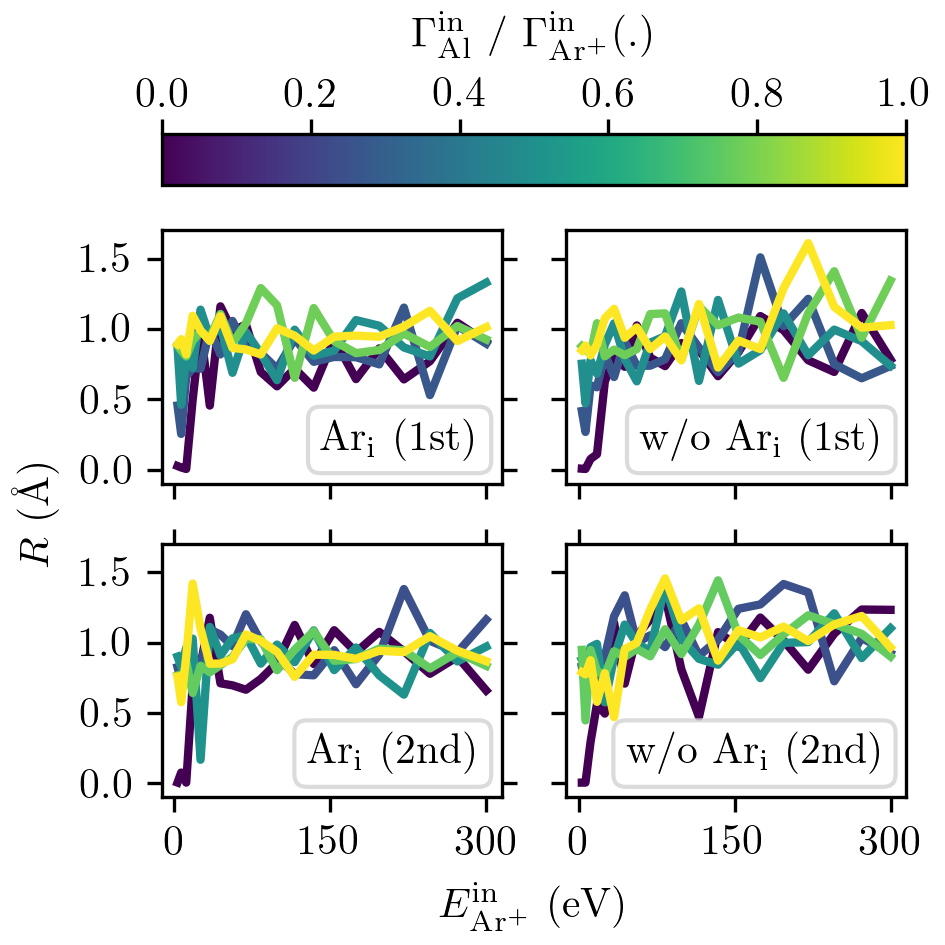}
\caption{The root-mean-squared roughness $R$ is displayed as a function of the Ar ion energy $E_\mathrm{Ar^+}^\mathrm{in}$ and Al/Ar flux ratio $\Gamma_\mathrm{Al}^\mathrm{in}/\Gamma_\mathrm{Ar^+}^\mathrm{in}$ for the 1st and 2nd set of impinging particles. Implanted Ar atoms are neglected.}
\label{fig:RMS}
\end{figure}

The root-mean-squared roughness $R$ is shown in Figure~\ref{fig:RMS}. It is approximately constant for any herein considered combination of process parameter. The evolution of this quantity is however anyhow restricted by the surfaces' lateral dimensions.

\begin{figure}[t]
\subfloat[Rings with 3 nodes ($n=3$).\label{sfig:ringRc3}]{%
    \includegraphics[width=8cm]{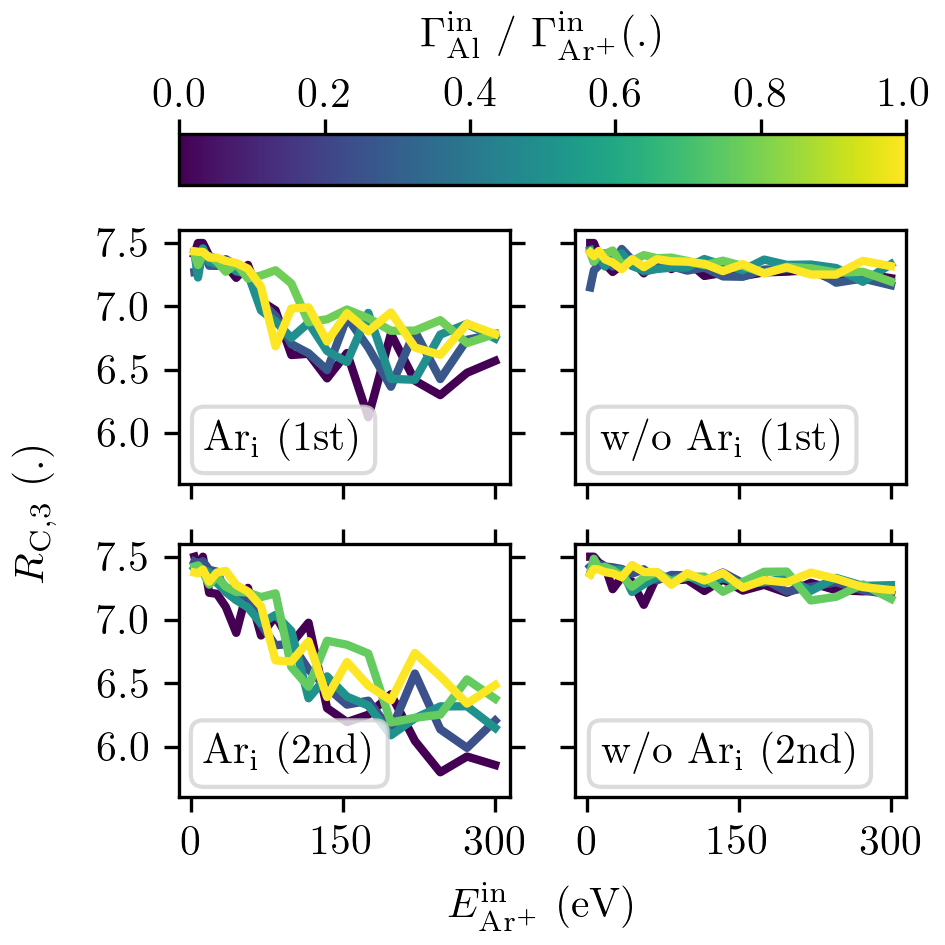}
}\hfill
\subfloat[Rings with 4 nodes ($n=4$).\label{sfig:ringRc4}]{%
    \includegraphics[width=8cm]{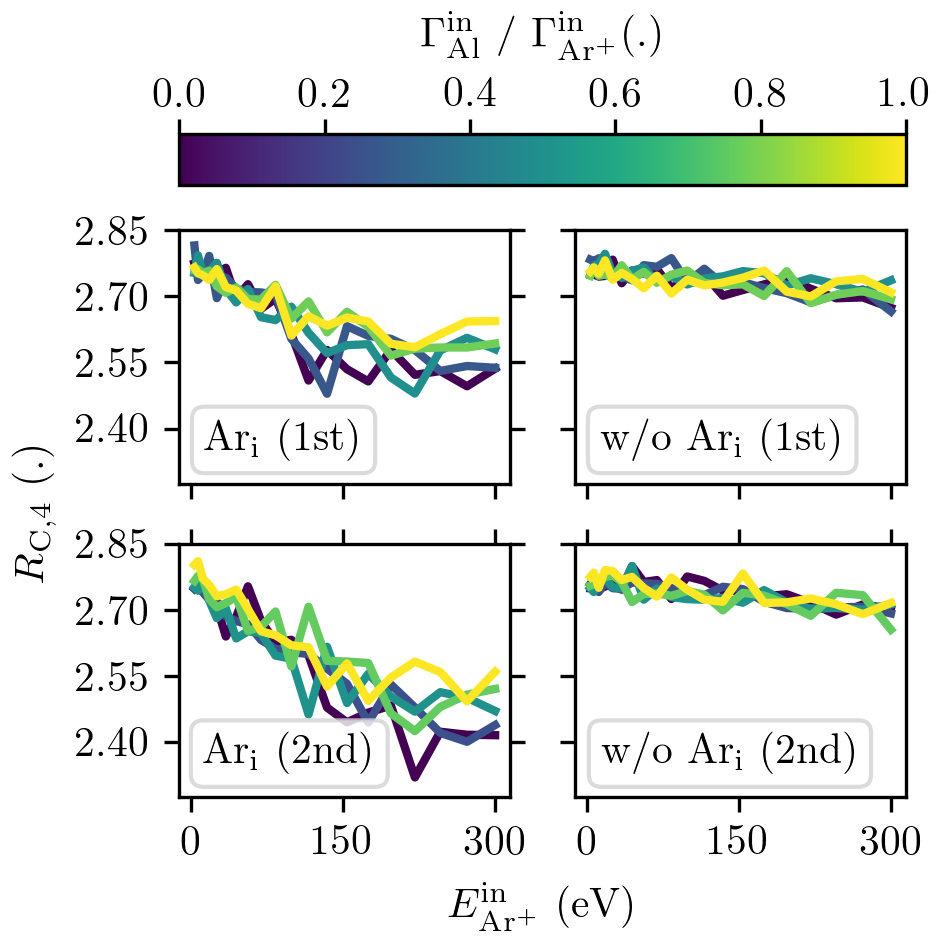}
}\hfill
\subfloat[Rings with 5 nodes ($n=5$).\label{sfig:ringRc5}]{%
    \includegraphics[width=8cm]{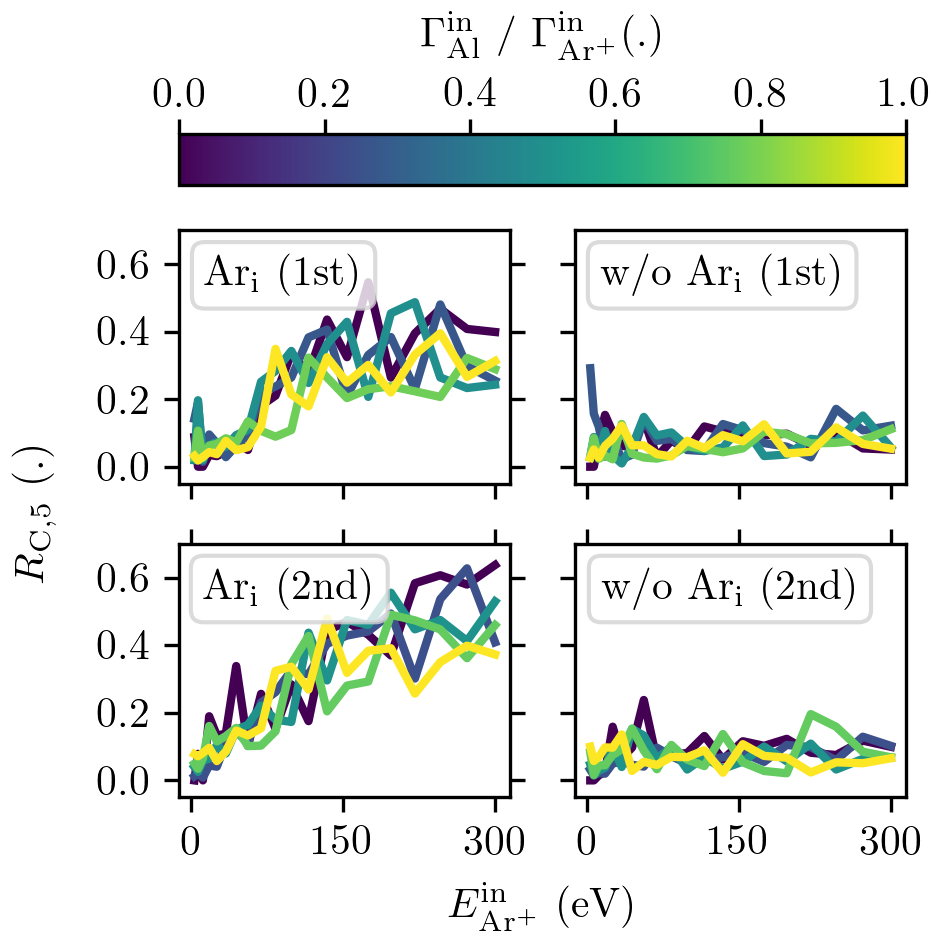}
}\hfill
\subfloat[Rings with 6 nodes ($n=6$).\label{sfig:ringRc6}]{%
    \includegraphics[width=8cm]{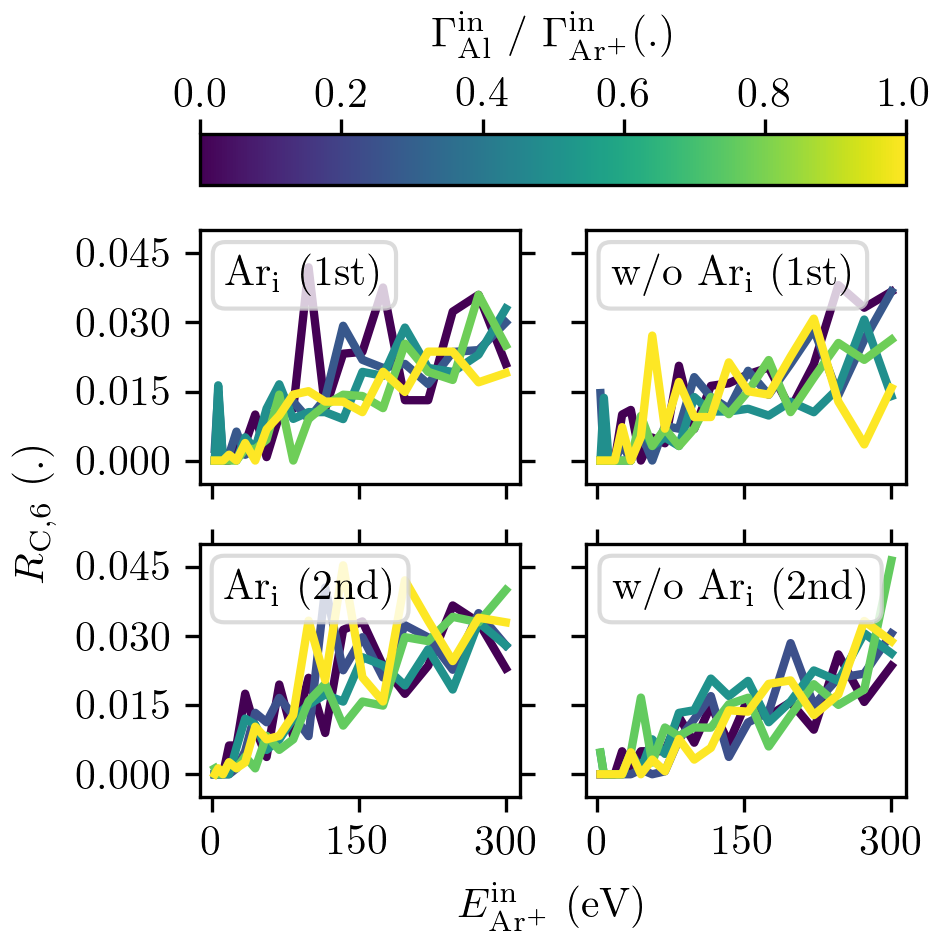}
}
\caption{The proportion of rings with $n$ nodes per atom $R_{\mathrm{C},n}$ is plotted against the Ar ion energy and Al/Ar flux ratio for the 1st and 2nd set as well as with and without implanted Ar atoms.}
\label{fig:ringRc}
\end{figure}

The description of the defect structure by means of the connectivity profile (ring statistics) is used in this work to support visual findings on the surface slabs' structural changes (i.e., vacancies, interstitials, Ar clusters, surface reconstructions). These require the consideration of rings with 5 and 6 nodes, while rings with 3 and 4 nodes are already sufficient to characterize the ideal fcc bulk lattice and Al(001) surface structure. Hence, the proportion of atoms at the origin of at least one ring with 3 or 4 nodes $P_{\mathrm{N},3}$ and $P_{\mathrm{N},4}$ equal approximately 1 for any herein considered combination of process parameters. The number of rings with 3 and 4 nodes normalized to the number of atoms $R_{\mathrm{C},3}$ and $R_{\mathrm{C},4}$ is slightly decreased for higher ion energies due to the accumulation of vacancies in either case, which is shown in Figures~\ref{sfig:ringRc3} and \ref{sfig:ringRc4}, respectively. This trend is amplified when incorporated Ar atoms are taken into account due to the deviations from the ideal fcc lattice structure, so far discussed in this section. 

\begin{figure}[t]
\subfloat[Rings with 5 nodes ($n=5$).\label{sfig:ringPn5}]{%
    \includegraphics[width=8cm]{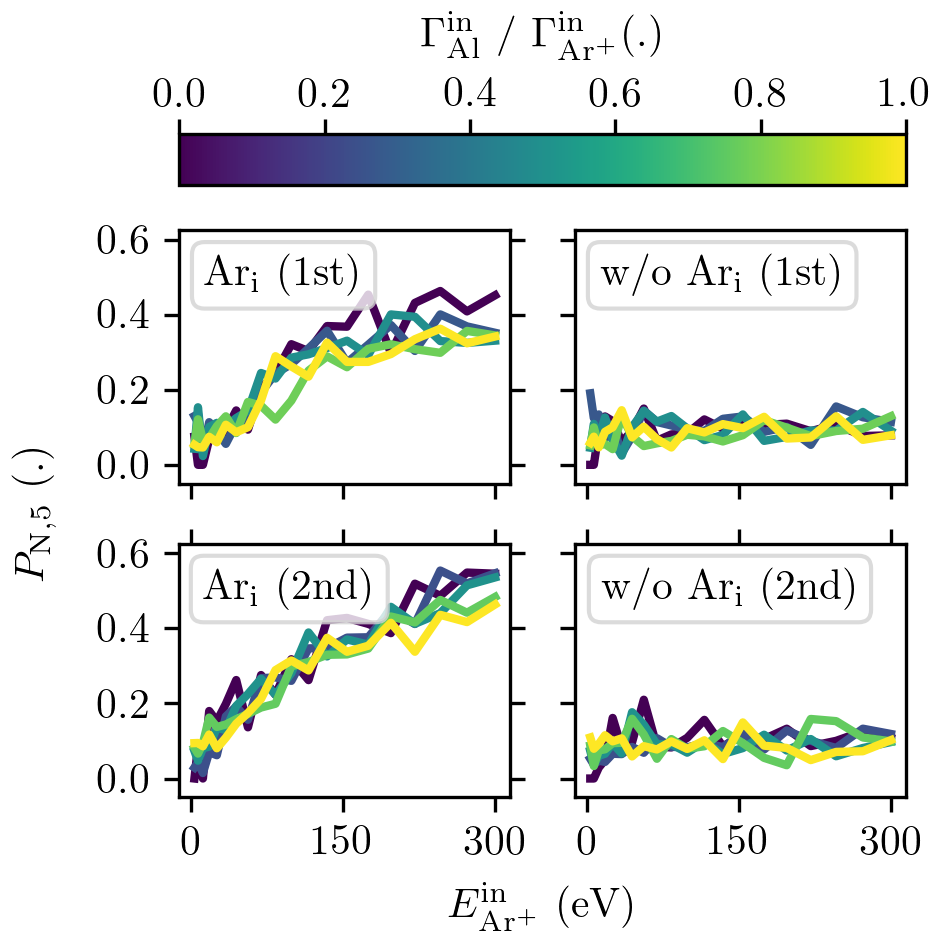}
}\hfill
\subfloat[Rings with 6 nodes ($n=6$). \label{sfig:ringPn6}]{%
    \includegraphics[width=8cm]{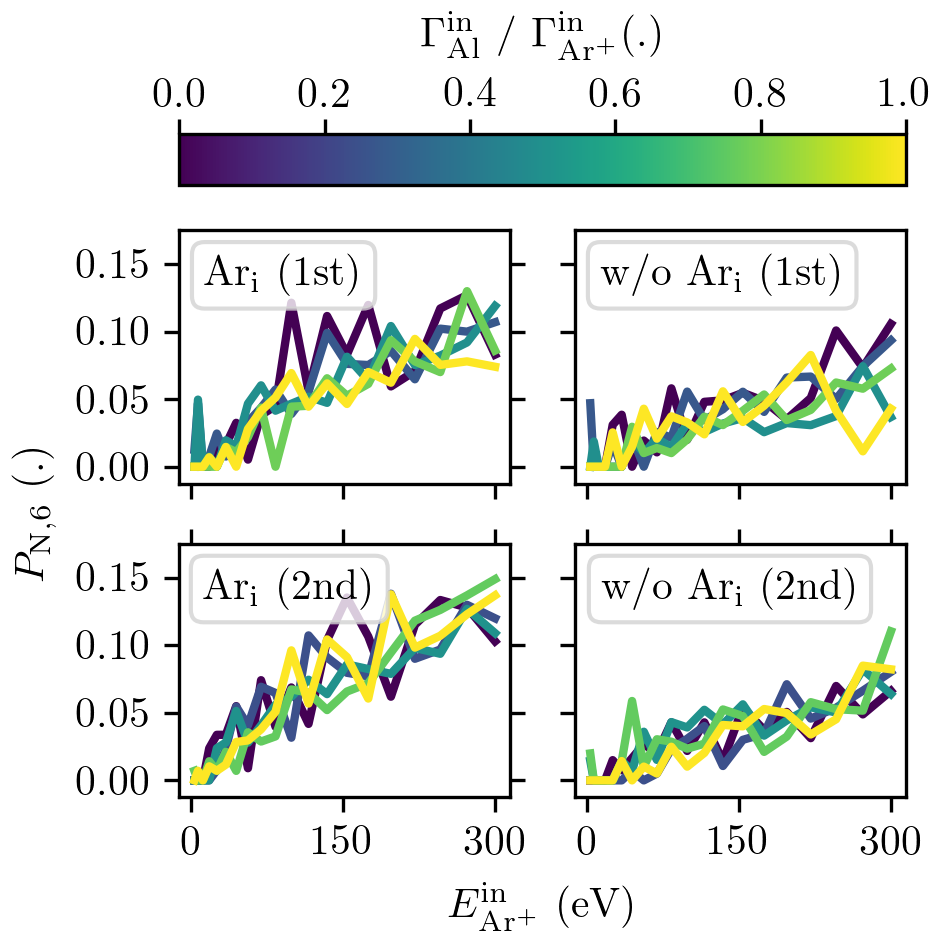}
}
\caption{The proportion of atoms at the origin of at least one ring with $n$ nodes $P_{\mathrm{N},n}$ is displayed as a function of the Ar ion energy $E_\mathrm{Ar^+}^\mathrm{in}$ and Al/Ar flux ratio $\Gamma_\mathrm{Al}^\mathrm{in}/\Gamma_\mathrm{Ar^+}^\mathrm{in}$ for the 1st and 2nd set of impinging particles as well as with and without implanted Ar atoms.}
\label{fig:ringPn}
\end{figure}

Surface reconstruction and forward sputtered (peened) Al atoms (interstitials) can be related to the rings with 5 nodes by means of $R_{\mathrm{C},5}$ and $P_{\mathrm{N},5}$, which are shown in Figure~\ref{sfig:ringRc5} and Figure~\ref{sfig:ringPn5}, respectively. Up to 55.43 \% of the Al atoms are at the origin of at least one ring with 5 nodes when ion energies of 300 eV are considered. Hence, more than half of the atoms in the surface slabs are part of at least minor structural changes. The lack of distorted surface layers or Al interstitials in case of neglected Ar implantation, therefore, results in a diminishing, process independent number of rings with 5 nodes.

Rings with 6 nodes are caused by Al vacancies and their amount is increased by implanted Ar atoms, which also occupy vacant Al lattice sites (if considered). Hence, greater ion energies lead to greater $R_{\mathrm{C},6}$ and $P_{\mathrm{N},6}$ in cases with or without implanted Ar atoms, which is shown in Figure~\ref{sfig:ringRc6} and Figure~\ref{sfig:ringPn6}. As expected, overall the ion energy dependencies of the described defect structure (connectivity profile) resemble the ones of the biaxial stresses, which are shown in Figure~\ref{sfig:bistress}. 


\section{Conclusion}
\label{sec:conclusion}

The sputter yield of Ar$^+$ ions bombarding Al(001) surfaces is found to be insensitive to the implantation of Ar atoms due to a balance of having additional recoil centers but also disturbing the uppermost surface layers' lattice structure. Ar atoms are distributed apart from each other within the surface slabs when the ion energies are below approximately 100 eV, occupying vacant Al lattice sites. Surrounding Al atoms are pushed away to keep an averaged distance of 3 \r A to the Ar atoms. For ion energies that exceed 100 eV the formation of Ar clusters is observed due to the deeper implantation, increased Ar concentration as well as collision cascade induced structural reordering. The outgassing of these Ar clusters is eased when the number of sputtered Al atoms exceeds the number of built-in Al atoms per impinging Ar ion $\left(\Gamma_\mathrm{Al}^\mathrm{in}/\Gamma_\mathrm{Ar^+}^\mathrm{in}<\Gamma_\mathrm{Al}^\mathrm{out}/\Gamma_\mathrm{Ar^+}^\mathrm{in}\right)$, due to the gradual removal of the enclosing surface layers. Incorporated Ar atoms interrupt the evolution of the collision cascades and distribute their momenta more evenly, which in combination with the induced stress and strain eventually hinders the formation of Al vacancies. Simultaneously they stabilize forward sputtered (peened) Al atoms (interstitials) by balancing the tensile surface stress. The effect of peened Al atoms on compressive thin film stresses may dominate the Ar atoms' contribution. But if it were not for implanted Ar atoms Al Frenkel pairs would readily diffuse, rearrange and recombine. This would result in thin films under even more tensile stresses the greater the ion energy. Ar atoms, Al vacancies and Al interstitials are found to balance each other with respect to the mass density, which is approximately constant unless larger Ar cluster form. These findings are supported by an analysis of the defect structure by means of a connectivity profile (ring statistics). Rings with 3 and 4 nodes per atom, which are sufficient to describe the ideal fcc lattice structure, are replaced by rings with 5 nodes (reconstructed surface layer and interstitials) and 6 nodes (vacancies and Ar atoms on vacant Al sites). This trend is more distinct with greater ion energies. An expected correlation of the defect structure with other surface properties (e.g., biaxial stress) can be identified.

Different doses complicate the interpretation of the simulation results, so that future investigations of similar kinds may rather set up case studies with a constant dose or impinging Ar rate (estimated process time) instead of a constant number of impinging particles. The lateral dimensions of the surface slab are limited by computational costs and, hence, fidelity, but ultimately do now allow for significant size effects. It is for instance assumed that the evolution of the surface roughness is limited by the considered surface areas. 

The results ultimately encourage the explicit consideration of implantation of even small Ar concentrations when studying growth or sputtering processes (by means of molecular dynamics simulations). The assumption that Ar atoms either outgas anyhow or do not significantly influence the ongoing dynamics is found to be invalid. 

\section*{Acknowledgement}

Funded by the Deutsche Forschungsgemeinschaft (DFG, German Research Foundation) – Project-ID 138690629 (TRR 87) and Project-ID 434434223 (SFB 1461).

\section*{ORCID iDs}
\noindent
T. Gergs: \url{https://orcid.org/0000-0001-5041-2941}\\
T. Mussenbrock: \url{https://orcid.org/0000-0001-6445-4990} \\
J. Trieschmann: \url{https://orcid.org/0000-0001-9136-8019}


\bibliography{./references.bib}

\end{document}